\documentclass[aip,pop,reprint]{revtex4-2}
\usepackage{graphicx}
\usepackage{amsmath}
\usepackage{mathrsfs}
\usepackage{array}
\usepackage[breaklinks=true,colorlinks=true,linkcolor=blue,urlcolor=blue,citecolor=blue]{hyperref}
\graphicspath{{../figures/}}

\begin{document}

\title{Characterization of fast magnetosonic waves driven by compact toroid plasma injection along a magnetic field} 
\author{F. Chu}
\email[]{fchu@lanl.gov}
\affiliation{Los Alamos National Laboratory, Los Alamos, New Mexico 87545, USA}
\author{S. J. Langendorf}
\affiliation{Los Alamos National Laboratory, Los Alamos, New Mexico 87545, USA}
\author{J. Olson}
\affiliation{Department of Physics, University of Wisconsin-Madison, Madison, Wisconsin 53706, USA}
\author{T. Byvank}
\affiliation{Los Alamos National Laboratory, Los Alamos, New Mexico 87545, USA}
\author{D. A. Endrizzi}
\affiliation{Department of Physics, University of Wisconsin-Madison, Madison, Wisconsin 53706, USA}
\author{A. L. LaJoie}
\affiliation{Department of Electrical and Computer Engineering, University of New Mexico, Albuquerque, New Mexico 87131, USA}
\affiliation{Los Alamos National Laboratory, Los Alamos, New Mexico 87545, USA}
\author{K. J. McCollam}
\affiliation{Department of Physics, University of Wisconsin-Madison, Madison, Wisconsin 53706, USA}
\author{C. B. Forest}
\affiliation{Department of Physics, University of Wisconsin-Madison, Madison, Wisconsin 53706, USA}
\date{\today}

\begin{abstract}
\rightskip.5in
Magnetosonic waves are low-frequency, linearly polarized magnetohydrodynamic (MHD) waves commonly found in space, responsible for many well-known features, such as heating of the solar corona. In this work, we report observations of interesting wave signatures driven by injecting compact toroid (CT) plasmas into a static Helmholtz magnetic field at the Big Red Ball (BRB) Facility at Wisconsin Plasma Physics Laboratory (WiPPL). By comparing the experimental results with the MHD theory, we identify that these waves are the fast magnetosonic modes propagating perpendicular to the background magnetic field. Additionally, we further investigate how the background field, preapplied poloidal magnetic flux in the CT injector, and the coarse grid placed in the chamber affect the characteristics of the waves. Since this experiment is part of an ongoing effort of creating a target plasma with tangled magnetic fields as a novel fusion fuel for magneto-inertial fusion (MIF), our current results could shed light on future possible paths of forming such a target for MIF.
\end{abstract}

\maketitle

\section{Introduction}
\label{sec:intro}

Magnetosonic waves are a type of low-frequency, linearly polarized magnetohydrodynamic (MHD) waves that can be excited in any electrically conducting fluid permeated by a magnetic field. Two magnetosonic modes can be derived from the MHD theory -- the fast and slow waves with both longitudinal (i.e., sound wave) and transverse (i.e., electromagnetic wave) components \cite{boardsen_survey_2016, gurnett_introduction_2017}. Magnetosonic waves are commonly observed in space \cite{wang_doppler_2002, ofman_slow_2012, yuan_alfven_2021}, and have gathered significant attention in recent years as they play important roles in a variety of astrophysical phenomena. For example, Pekünlü et al.~conducted a theoretical investigation on fast magnetosonic waves in solar coronal loops and showed that these waves were possible agents for heating the solar corona through wave energy dissipation \cite{pekunlu_solar_2001}. Using Mars Atmosphere and Volatile Evolution (MAVEN) spacecraft data, Fowler et al.~reported that large-amplitude magnetosonic waves were observed within the magnetosheath at Mars, driving significant ion and electron heating in the Martian dayside ionosphere \cite{fowler_maven_2018}.

By far most of the experimental studies on magnetosonic waves have been carried out in space. However, the \textit{in situ} spacecraft measurements at only a single or a few points have generally suffered from limitations of insufficient spatial resolution and uncontrolled conditions \cite{howes_laboratory_2018}. The laboratory simulation of space plasmas, on the other hand, enables detailed investigations of underlying plasma physics processes \cite{howes_toward_2012, schroeder_laboratory_2021, valenzuela-villaseca_characterization_2023} with many-point measurements, reproducible environments, and a suite of state-of-the-art diagnostics \cite{schroeder_direct_2016, chu_determining_2019, chu_measurement_2019, milhone_spectrometer_2019, shi_multi-dimensional_2023}. The Big Red Ball (BRB) Facility at the Wisconsin Plasma Physics Laboratory (WiPPL) \cite{cooper_madison_2014, forest_wisconsin_2015} is designed to study a range of fundamental astrophysical questions as well as geometries that mimic astrophysical systems, providing a unique capability to investigate the fast magnetosonic waves in a laboratory.

In this paper, we report an experiment of driving the fast magnetosonic waves through injection of compact toroid (CT) plasma \cite{degnan_compact_1993, bellan_spheromaks_2000, slutz_pulsed-power-driven_2010} along a magnetic field in BRB. This experiment is part of an ongoing exploration of forming a target plasma with tangled magnetic fields in a laboratory setting as a novel fusion fuel for magneto-inertial fusion (MIF), also known as magnetized target fusion (MTF) \cite{kirkpatrick_magnetized_1995, lindemuth_fundamental_2009, hsu_magnetized_2019}. In this approach, the target plasma is quasi-adiabatically compressed and heated by heavy imploding shell or ``liner'', with the goal of briefly attaining thermonuclear burn conditions. As the electron heat conduction goes predominantly along the field lines, the randomized magnetic fields in the target can provide very long connection lengths between the core and the liner surface, therefore effectively reduce heat loss from the fuel plasma to the colder liner \cite{ryutov_adiabatic_2009}. 

In the experiment performed in BRB, we intend to create a target plasma with tangled fields through turbulence by colliding the CT with a coarse conducting grid placed in the chamber. During this process, however, we often observe interesting wave signatures appearing in the direction along the background magnetic field. The main scope of this paper is to investigate the nature of these waves. We first identify the wave modes by comparing the experimental data with MHD theory, and then study how the background field, preapplied poloidal magnetic flux in the CT injector, and the coarse grid affect the characteristics of the waves. At the end of the paper, we will also discuss our future possible path of forming a target plasma with tangled fields for MIF based on our current results.

The paper is organized as follows: Sec.~\ref{sec:background} presents a background on fast magnetosonic waves, Sec.~\ref{sec:exp} gives a description of the experimental setup and the coaxial gun used to create CT plasmas, Sec.~\ref{sec:results} presents the experimental results, Sec.~\ref{sec:discussion} discusses wave driving mechanisms, target plasma formation for MIF, and possible path forward, and Sec.~\ref{sec:summary} provides a summary.



\section{Background on Fast Magnetosonic Waves}
\label{sec:background}

To understand the basic properties of the magnetosonic waves, in this section, we consider a simple case of small-amplitude waves in a homogeneous ideal (infinitely conducting) MHD fluid. The closed set of ideal MHD equations, including the continuity equation, momentum equation, induction equation, and adiabatic equation of state, are \cite{bellan_fundamentals_2008, gurnett_introduction_2017}
\begin{gather}
\label{eq:MHD1}
\frac{\partial \rho }{\partial t}+\nabla\cdot \left ( \rho \mathbf{u} \right )=0, \\
\label{eq:MHD2}
\rho \frac{\partial\mathbf{u} }{\partial t}+\rho (\mathbf{u}\cdot \nabla) \mathbf{u}=-\nabla\left ( P+\frac{B^2}{2\mu _0} \right )+\frac{(\mathbf{B}\cdot \nabla)\mathbf{B}}{\mu _0},\\
\label{eq:MHD3}
\frac{\partial \mathbf{B}}{\partial t}=\nabla \times (\mathbf{u} \times \mathbf{B}),\\
\label{eq:MHD4}
\frac{\mathrm{d}}{\mathrm{d}t}\left ( \frac{P}{\rho ^\gamma } \right )=0,
\end{gather}
where $\rho$ is the mass density, $\mathbf{u}$ is the fluid velocity, $\mathbf{B}$ is the magnetic field, $P$ is the pressure (isotropic pressure approximation), $\mu_0$ is the vacuum permeability, and $\gamma$ is the adiabatic index (ratio of specific heats). To obtain wave equations, the above system of equations is linearized by assuming that $\rho$, $\mathbf{u}$, $\mathbf{B}$, and $P$ are the sum of a spatially uniform time-independent equilibrium quantity (denoted with the subscript ``0'') plus a small first-order perturbation (denoted with the subscript ``1''), i.e., $\rho =\rho _0+\rho _1$, $\mathbf{u}=\mathbf{u}_1$, $\mathbf{B}=\mathbf{B}_0+\mathbf{B}_1$, and $P =P _0+P _1$.

\begin{figure}[h]
\begin{center}
\includegraphics[width=3.37in]{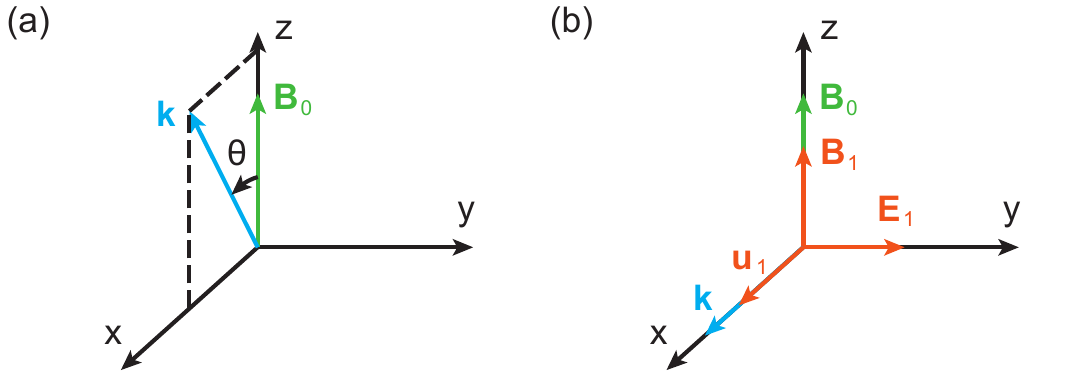}
\caption{(a) Setup of the coordinate system for MHD wave analysis. The background magnetic field $\mathbf{B}_0$ is along the $z$-axis and the wave vector $\mathbf{k}$ lies in the $x$-$z$ plane. The angle subtended between $\mathbf{k}$ and $\mathbf{B}_0$ is denoted by $\theta$. (b) The eigenvectors for the fast magnetosonic mode in a special case where the direction of the wave propagation is perpendicular to the background magnetic field ($\theta=\pi/2$). Perturbations in the magnetic field, electric field, and plasma motion are denoted by $\mathbf{B}_1$, $\mathbf{E}_1$, and $\mathbf{u}_1$, respectively.}
\label{fig:wave}
\end{center}
\end{figure}



Without loss of generality, as shown in Fig.~\ref{fig:wave}(a), we can assume that the equilibrium magnetic field $\mathbf{B}_0=(0,0,B_0)$ is directed along the $z$-axis and the wave vector $\mathbf{k}=(k\sin\theta ,0, k\cos\theta )$ lies in the $x$-$z$ plane, where $\theta$ is the angle between $\mathbf{B}_0$ and $\mathbf{k}$. By Fourier-analyzing Eqs.~(\ref{eq:MHD1})--(\ref{eq:MHD4}) and making the operator substitutions ($\nabla \rightarrow \mathrm{i}\mathbf{k}$ and $\partial/\partial t \rightarrow -\mathrm{i}\omega$), we can obtain the eigenvalue equation 
\begin{gather}
\begin{pmatrix}
v_p^2-V_S^2\sin^2\theta -V_A^2 & 0 &-V_S^2\sin\theta \cos\theta \\ 
0 & v_p^2-V_A^2\cos^2\theta & 0\\ 
 -V_S^2\sin\theta \cos\theta & 0 & v_p^2-V_S^2\cos^2\theta
\end{pmatrix}  \nonumber \\
\label{eq:eigenvalueequation}
\cdot \begin{pmatrix}
u_{x1}\\ 
u_{y1}\\ 
u_{z1}
\end{pmatrix}=0,
\end{gather}
where $v_p=\omega/k$ is the phase velocity, $V_S=\sqrt{\gamma P_0/\rho _0}$ is the sound speed, and $V_A=\sqrt{B^2_0/\mu_0 \rho _0}$ is the Alfvén speed. The above eigenvalue equation has non-trivial solutions for $\mathbf{u}_1$ if and only if the determinant of the matrix on the left side of Eq.~(\ref{eq:eigenvalueequation}) is zero,
which yields the dispersion relation
\begin{align}
\label{eq:dispersion}
D(\omega,k)=&\big ( v_p^2-V_A^2\cos^2\theta  \big ) \Big ( v_p^4-v_p^2 \big ( V_A^2+V_S^2 \big) \nonumber \\
&+V_A^2V_S^2\cos^2\theta \Big)=0.
\end{align} 

It can be shown that the dispersion relation has three independent roots, corresponding to the three different wave modes that can propagate through an MHD plasma
\begin{gather}
\label{eq:roots1}
v_p^2=V_A^2\cos^2\theta, \\
\label{eq:roots2}
v_{p\pm}^2=\frac{1}{2} \bigg( V_A^2+V_S^2  \pm \sqrt{ \big( V_A^2-V_S^2 \big)^2 + 4V_A^2V_S^2\sin^2\theta }\bigg).
\end{gather} 
The roots in Eqs.~(\ref{eq:roots1}) and (\ref{eq:roots2}) are called the shear Alfvén mode, fast magnetosonic mode, and slow magnetosonic mode, respectively. Note that $v_{p+} \geq v_{p-}$. The fast magnetosonic mode in Eq.~(\ref{eq:roots2}) can also be considered as the compressional Alfvén mode ($V_S \rightarrow 0$) modified by a non-zero plasma pressure \cite{fitzpatrick_plasma_2014}.

\begin{figure*}
\begin{center}
\includegraphics[width=6.69in]{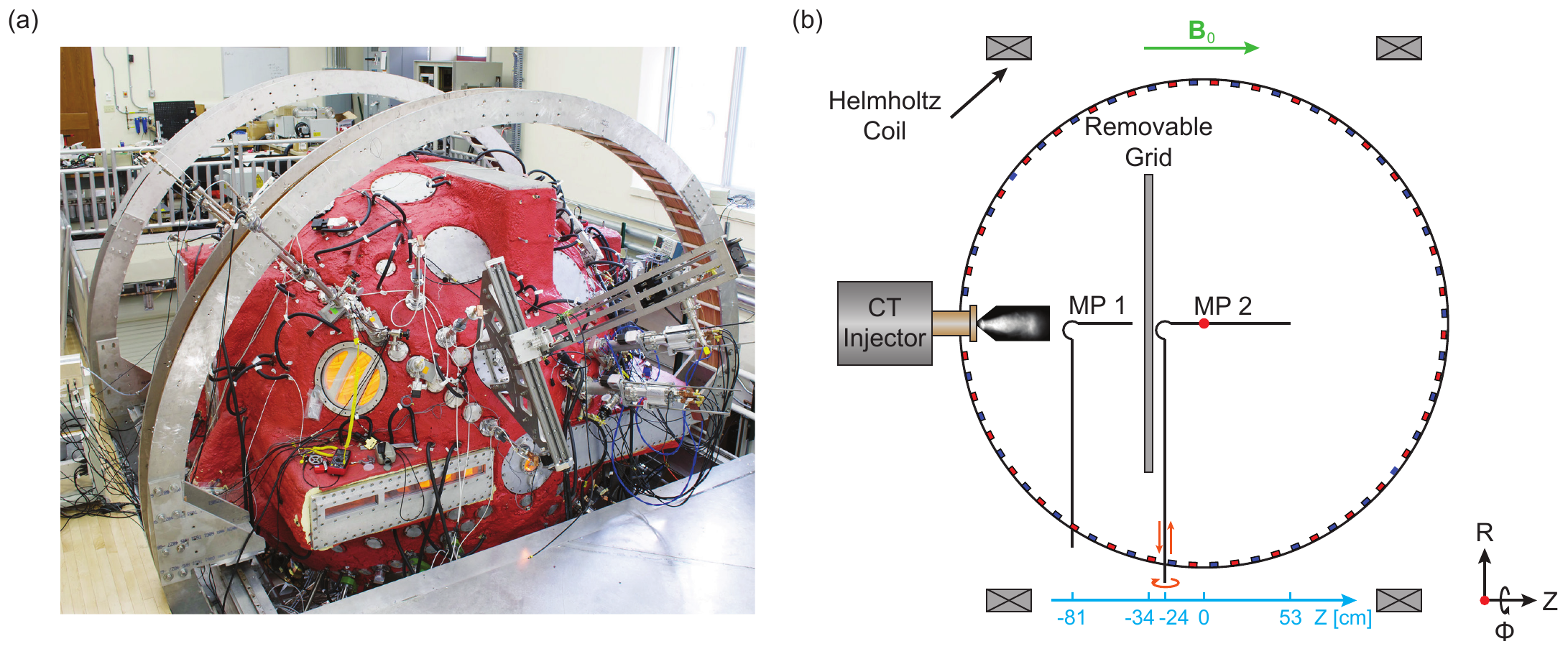}
\caption{Image of the BRB (a) and poloidal cross-section of the experimental setup (b). The hydrogen CT plasma is injected into the vacuum chamber by a coaxial plasma gun mounted on the southern magnetic pole of the vessel. The chamber coordinates in this paper are described in a cylindrical coordinate system $R$-$\phi$-$Z$, with the origin (represented by the red dot) located in the center of the chamber. A removable coarse conducting grid used to interact with the CT plasmas is placed at $Z=-34$ cm. The topology or spatiotemporal structure of the magnetic field is measured using two linear arrays of three-axis magnetic probes ($\dot{B}$ probes). The magnetic probe array (MP1) in front of the grid is located between $Z=-81$ and $-44$ cm and the one (MP2) behind the grid covers positions from $-24$ to $53$ cm in $Z$. The probe arrays can also be rotated $90^\circ$ around their primary shafts to measure the magnetic field topology in the $R$ direction, as illustrated by the orange arrows.}
\label{fig:exp}
\end{center}
\end{figure*}

In a special case where $\theta=\pi/2$, the only non-trivial solution of the dispersion relation in Eq.~(\ref{eq:dispersion}) is the fast magnetosonic mode with the wave vector $\mathbf{k}=(k_x ,0, 0 )$ and phase velocity $v_p^2=V_A^2+V_S^2$. The eigenvectors for these waves are $\mathbf{u}_1=(u_{x1},0,0)$, $\mathbf{B}_1=(0,0,B_{z1})$, and $\mathbf{E}_1=(0,E_{y1},0)$, as shown in Fig.~\ref{fig:wave}(b). Since $\rho_1=(\rho_0/\omega)\mathbf{k} \cdot \mathbf{u}_1=\rho_0(u_{x1}/v_p)$, these eigenvectors suggest that this wave mode is associated with non-zero perturbations in the plasma density and pressure (compressible), sharing features of both a sound wave and an electromagnetic wave \cite{gurnett_introduction_2017}. In addition, as seen in Fig.~\ref{fig:wave}(b), these waves involve magnetic field perturbation parallel with the background magnetic field $\mathbf{B}_0$ ($\mathbf{B}_1 \parallel \mathbf{B}_0$) and plasma motion perpendicular to $\mathbf{B}_0$ ($\mathbf{u}_1 \perp \mathbf{B}_0$). 

We will show later in Sec.~\ref{sec:results} that the waves observed in the experiment resulting from interaction between CT plasmas and the background magnetic fields in BRB can be identified as the fast magnetosonic mode with $\theta=\pi/2$, as described in Fig.~\ref{fig:wave}(b). Please note that even though the CT propagates downstream parallel with $\mathbf{B}_0$ at a velocity $\mathbf{u}_0$ in the experiment, this equilibrium fluid velocity will not alter the solution of the MHD equations (\ref{eq:MHD1})--(\ref{eq:MHD4}) for $\theta=\pi/2$ shown above, due to the geometry between $\mathbf{u}_0$, $\mathbf{u}_1$, $\mathbf{k}$, and $\mathbf{B}_1$ ($\mathbf{u}_0 \perp \mathbf{u}_1$, $\mathbf{u}_0 \perp \mathbf{k}$, and $\mathbf{u}_0 \parallel \mathbf{B}_1$).

\section{Experimental Setup}
\label{sec:exp}

The experiment setup of forming the fast magnetosonic waves in BRB is depicted in Fig.~\ref{fig:exp}. The BRB is a 3-meter-diameter spherical chamber with multicusp magnetic plasma confinement. The cusp field is generated by an array of permanent magnets with alternating polarity and highly concentrated at the chamber wall, leaving the plasma in the core unmagnetized. The BRB is also equipped with an external 3-meter-diameter Helmholtz coil pair that provides a near-uniform axial magnetic field pointing from the southern magnetic pole to north (positive direction) throughout the plasma volume of up to 275 G. This background field helps prevent the CT plasmas from expanding too much as they travel inside the chamber \cite{byvank_formation_2021}, which can be seen later in Fig.~\ref{fig:idwave}. In the rest of the paper, the coordinates in the chamber are described in a cylindrical coordinate system $R$-$\phi$-$Z$ as illustrated in Fig.~\ref{fig:exp}(b), with the origin located in the center of the chamber.

\begin{figure}
\begin{center}
\includegraphics[width=3.37in]{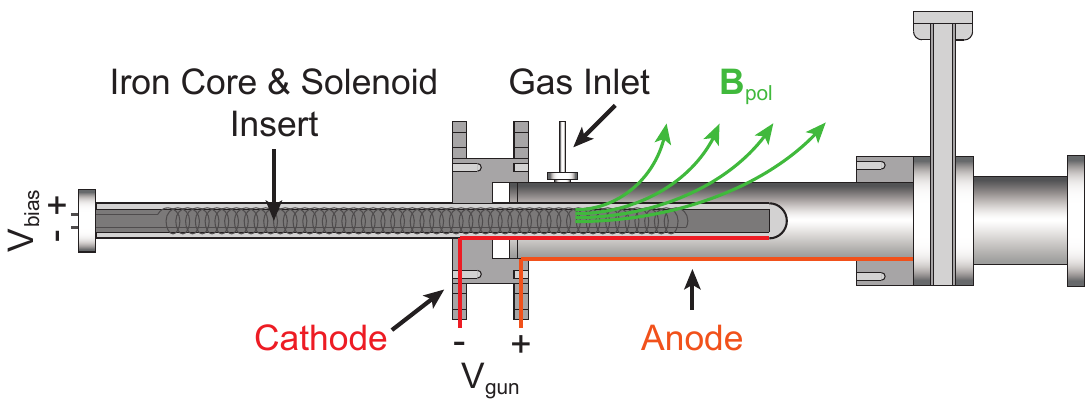}
\caption{Schematic of the CT injector used in the experiment. When the CT injector is fired, a hydrogen gas puff is first preionized between the anode and cathode, and then accelerated into the vacuum chamber along the $Z$ direction. The iron core and solenoid inserted into the inner electrode generate a poloidal bias flux $\Psi_\textup{gun}$. The magnitude and polarity of $\Psi_\textup{gun}$ are controlled by the DC voltage $V_\textup{bias}$ across the solenoid (or $I_\textup{bias}$ in the solenoid).}
\label{fig:gun}
\end{center}
\end{figure}


The hydrogen CT plasmas used to interact with the Helmholtz field are created by a coaxial plasma gun \cite{matsumoto_development_2016} mounted on the southern pole of the chamber. The diagram of the CT injector is illustrated in Fig.~\ref{fig:gun}. To produce CTs, a bias magnetic flux $\Psi_\textup{gun}$ linking the two electrodes is required for magnetizing the plasma. The magnetic field lines can then reconnect and detach from the electrodes as the plasma propagates out of the injector. In the experiment, an iron core surrounded by a copper winding of 4.1 cm diameter is inserted into the inner electrode and a poloidal bias flux of up to 0.4 mWb \cite{byvank_formation_2021} can therefore be established by supplying a DC current $I_\textup{bias}$ through the winding, as shown in Fig.~\ref{fig:gun}. In addition, the polarity of the bias magnetic flux $\Psi_\textup{gun}$ can be altered by switching the direction of the current in the winding. At the time when the CT injector is fired, a hydrogen gas puff is first preionized between the two coaxial electrodes and then accelerated axially (in parallel with the Helmholtz field) into the vacuum chamber at a velocity of $v_\textup{CT} \approx 70$ km/s. The CT plasmas are estimated to have a radius $r \approx 4$ cm, length $l \approx 10$ cm, electron and ion temperature $T_e \approx T_i \approx 30$ eV, density $n_e \approx n_i \approx 5 \times 10^{15}$ $\textup{cm}^{-3}$, and poloidal magnetic field $B \approx 2000$ G near the gun nozzle \cite{byvank_formation_2021}.

A removable coarse conducting grid of mesh size 10 $\times$ 10 cm, as shown in Fig.~\ref{fig:exp}, is placed at $Z = -34$ cm to interact with the CT. The topology or spatiotemporal structure of the magnetic field in the pregrid and postgrid region is measured using two linear arrays of three-axis magnetic probes ($\dot{B}$ probes), where the signal from the probe pickup coils is proportional to $\partial B/\partial t$. The probe array in front of the grid consists of 8 equally spaced probes located between $Z = -81$ and $-44$ cm, while the array behind the grid contains 15 equally spaced probes covering positions from $-24$ to $53$ cm in $Z$. A 2D structure of the magnetic field can therefore be obtained by scanning the two probe arrays in the radial direction. In addition, the probe arrays can be rotated $90^\circ$ around their primary shafts to measure the magnetic field topology in the $R$-direction, as shown in Fig.~\ref{fig:exp}(b). Electron densities and temperatures in the experiment are measured using a Langmuir probe with 16 closely spaced, individually biased tips, allowing the current-voltage ($I$-$V$) traces to be obtained with a 2 MHz resolution at the location of the probe \cite{olson_experimental_2016, byvank_formation_2021}.

\section{Experimental Results}
\label{sec:results}

We observe in the experiment that the CT plasmas have a radius $r \approx 30$ cm, length $l \approx 80$ cm, electron $T_e \approx 15$ eV, ion temperature $T_i \approx 30$ eV, and density $n_e \approx n_i \approx 1 \times 10^{13}$ $\textup{cm}^{-3}$ when they just reach the center of the vacuum chamber \cite{byvank_formation_2021}. The ion temperature is estimated from a previous measurement using ion Doppler spectroscopy \cite{langendorf_experimental_2018, chu_experimental_2023}, as described in Ref.~\onlinecite{byvank_formation_2021}. Based on the above plasma parameters, assuming a typical Helmholtz field $B_0=60$ G, we can estimate some important characteristic scales in the CT plasmas: ion cyclotron frequency $f_{ci} \approx 92$ kHz, electron cyclotron frequency $f_{ce} \approx 168$ MHz, plasma frequency $f_p \approx 28$ GHz, ion gyroradius $\rho_{ci} \approx 9$ cm, electron gyroradius $\rho_{ce} \approx 0.15$ cm, and Debye length $\lambda_D \ll 0.01$ cm. 


\subsection{Identification of the Wave Modes Observed in the Experiment}
\label{subsec:identification}

\begin{figure}
\begin{center}
\includegraphics[width=3.37in]{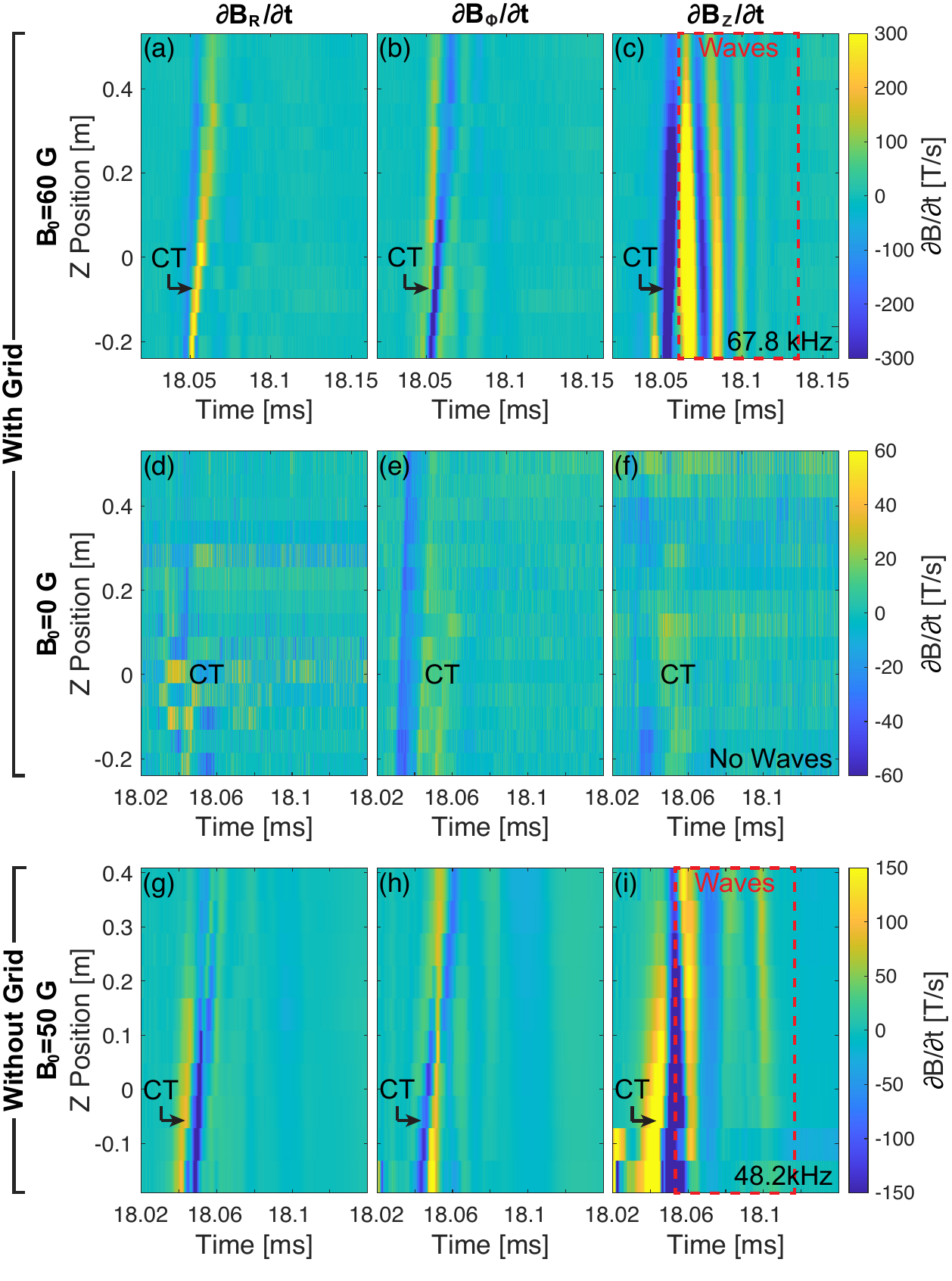}
\caption{Topology of the magnetic field $\partial B_R/\partial t$, $\partial B_\phi/\partial t$, $\partial B_Z/\partial t$ along the $Z$-axis near $R=0$ cm detected by probe array MP2. The data with the background Helmholtz field $B_0=60$ G are shown in (a)--(c) and $B_0=0$ G in (d)--(f). Both measurements are taken with the grid in the chamber. The time is referenced to the moment when the shot is executed in the control system. To confirm wether the waves are generated by the induced current in the grid, we repeat the shot without the grid and the data are shown in (g)--(i), where the background field $B_0=50$ G. The bias current in all cases is $I_\textup{bias}=3$~A. Waves with frequency 67.8 and 48.2 kHz are observed in (c) and (i), respectively, as shown in the red box. No waves are observed in (f).}
\label{fig:idwave}
\end{center}
\end{figure}

\begin{figure*}
\begin{center}
\includegraphics[width=4.33in]{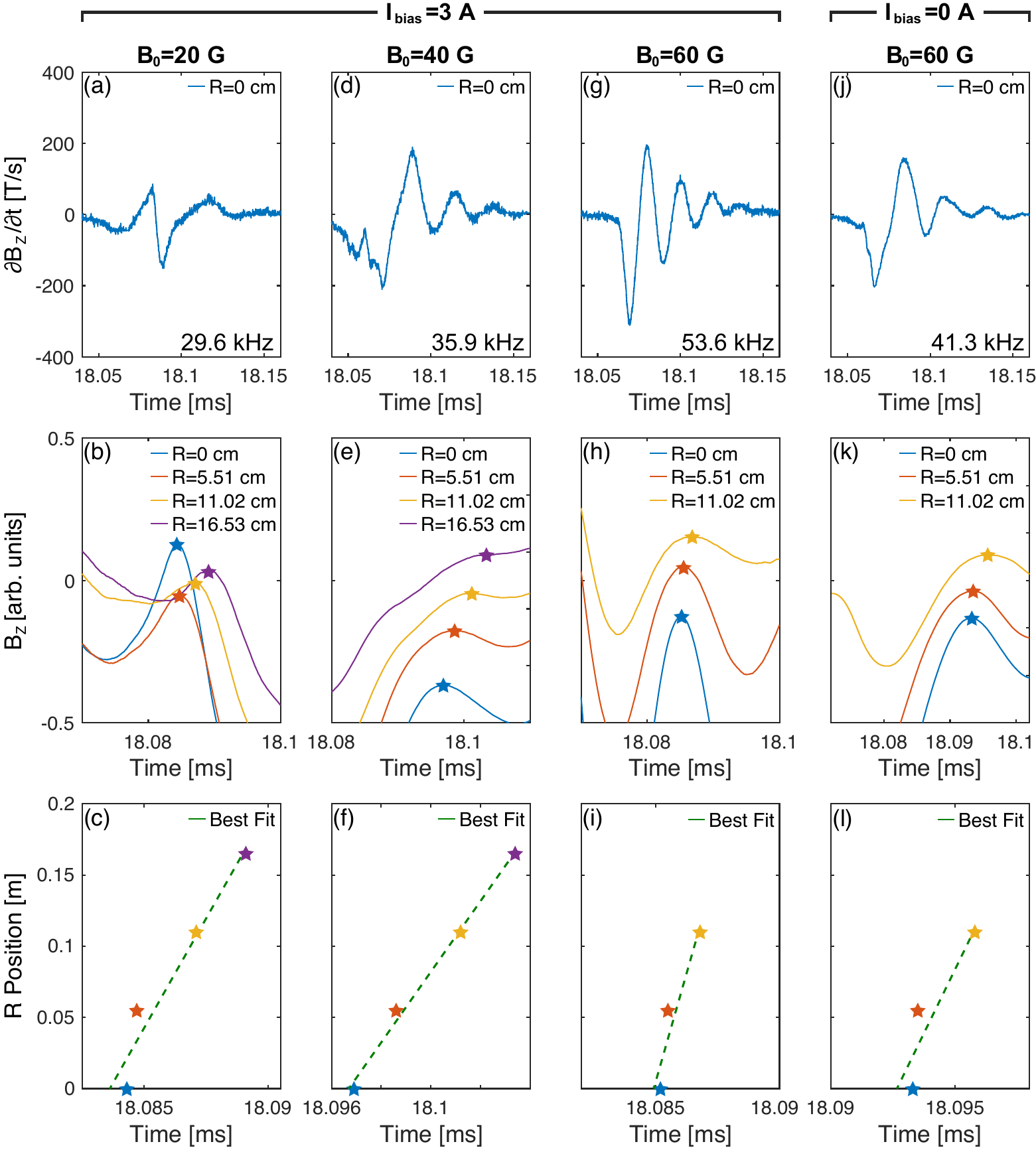}
\caption{Topology of the magnetic field $\partial B_Z/\partial t$ and $B_Z$ ($\partial B_Z/\partial t$ integrated over time) along the $R$-axis based on the measurements from probe array MP2. The data with the bias current $I_\textup{bias}=3$~A and background Helmholtz field $B_0=20$ G are shown in (a)--(c), $B_0=40$ G in (d)--(f), and $B_0=60$ G in (g)--(i). The data with $I_\textup{bias}=0$~A and $B_0=60$ G are shown in (j)--(l). In each $B_0$, the top panel shows $\partial B_Z/\partial t$ at $R=0$ cm to illustrate the waveform, while the middle panel shows $B_Z$ at multiple probe locations to demonstrate wave propagation along the $R$-axis. The bottom panel shows the time when peak of the waveform arrives at each probe location. The peak of each waveform is marked by a star. The experimental phase velocity $v_{p\text{-}\textup{exp}}$ can be determined from the linear fit of these data points. Note that (b), (e), (h), and (k) are zoomed in on time with the window size of $\Delta t=30$ $\mu$s.}
\label{fig:waveB}
\end{center}
\end{figure*}

We measure the topology of the magnetic field $\partial B/\partial t$ along the $Z$-axis near $R=0$ cm in the postgrid region (with the grid in the chamber) after the CT plasmas are injected into the chamber and the results are presented in Fig.~\ref{fig:idwave}. The time is referenced to the moment when the shot is executed in the control system. Figures~\ref{fig:idwave}(a)--\ref{fig:idwave}(c) show the shot fired when the background Helmholtz field $B_0=60$ G, while Figs.~\ref{fig:idwave}(d)--\ref{fig:idwave}(f) for $B_0=0$ G. In both cases, the bias current is kept at $I_\textup{bias}=3$ A, resulting in a poloidal bias flux $\Psi_\textup{gun}=0.4$ mWb. 


Two prominent features can be immediately noticed in Figs.~\ref{fig:idwave}(a)--\ref{fig:idwave}(c). The first is the CT plasma propagating along the probe array from ${\sim} 18.05$ to ${\sim} 18.08$ ms. After the CT travels through and expands in the chamber, a wave pattern with frequency $f_\textup{wave}=67.8 \textup{ kHz} \approx 0.74f_{ci}$ is detected by the probe in the $Z$-axis between ${\sim} 18.08$ and ${\sim} 18.13$ ms, as shown in the red box in Fig.~\ref{fig:idwave}(c). When this shot is repeated with $B_0=0$ G, however, no wave patterns are observed in the data except the initial CT plasma passing by the probe array, suggesting that the background magnetic field $B_0$ is required to support these waves. The signal in this shot is much weaker compared to $B_0=60$ G due to plasma expansion in the absence of the confinement from the Helmholtz field. 


We also fire additional shots with similar experimental settings without the grid to confirm wether the waves are generated by the induced current in the grid and the results are presented in Figs.~\ref{fig:idwave}(g)--\ref{fig:idwave}(i). It shows that the wave pattern can still be observed without the grid, which suggests that these waves are indeed driven by the interaction between the CT and the background magnetic field.

Since the wave frequency $f_\textup{wave}$ is smaller than $f_{ci}, f_{ce}, f_p$ and the CT plasma size is greater than $\rho_{ci}, \rho_{ce}, \lambda_D$, we can conclude that the wave modes observed in Fig.~\ref{fig:idwave}(c) are MHD waves. As we can see later in Sec.~\ref{subsec:HelmholtzB}, the wavelength also satisfies the condition $\lambda_\textup{wave}>\rho_{ci}, \rho_{ce}, \lambda_D$. In addition, the wave magnetic field $\mathbf{B}_1$ is mostly parallel with $\mathbf{B}_0$ along the $Z$-axis and the wave vector $\mathbf{k}$ is always perpendicular to $\mathbf{B}_1$ \cite{fitzpatrick_plasma_2014, gurnett_introduction_2017}, suggesting that $\mathbf{k} \perp \mathbf{B}_0$. Thus, according to the theory presented in Sec.~\ref{sec:background}, we can identify the waves observed in our experiment as the fast magnetosonic mode propagating perpendicular to the background magnetic field $\mathbf{B}_0$.

\subsection{Effects of the Background Magnetic Field $\mathbf{B}_0$}
\label{subsec:HelmholtzB}


Since the wave vector $\mathbf{k}$ points in the radial direction, the probe arrays need to be rotated $90^\circ$ from their original positions shown in Fig.~\ref{fig:exp}(b) to spatially resolve wave propagation along the $R$-axis. The results of such measurements obtained by probe array MP2 are shown in Fig.~\ref{fig:waveB}, where panels \ref{fig:waveB}(a)--\ref{fig:waveB}(c) are for the case $B_0=20$ G, \ref{fig:waveB}(d)--\ref{fig:waveB}(f) for $B_0=40$ G, and \ref{fig:waveB}(g)--\ref{fig:waveB}(i) for $B_0=60$ G. The bias current $I_\textup{bias}$ in these measurements is kept at 3 A. More specifically in each case, the top panel presents the raw $\partial B_Z/\partial t$ signal at $R=0$ cm to illustrate the waveform, while the middle panel shows $B_Z$ ($\partial B_Z/\partial t$ integrated over time) at multiple probe locations, where waves can be seen propagating along the $R$-axis. The bottom panel shows the time when peak of the waveform arrives at each probe location, used to calculate wave phase velocity. The measured wave properties ($f_\textup{wave}$, $\lambda_\textup{wave}$, and $v_{p\text{-}\textup{exp}}$) and comparison to the theoretical phase velocity $v_{p\text{-}\textup{theory}}$ at various $B_0$ are summarized in Table~\ref{tab:waves}.

It can be seen in Table~\ref{tab:waves} that the measured phase velocities $v_{p\text{-}\textup{exp}}$ are largely in agreement with the theoretical predictions $v_{p\text{-}\textup{theory}}$, given the error in evaluating the electron temperature and density ($T_e$ and $n_e$) in the experiment and simplified assumptions used in derivation of $v_{p\text{-}\textup{theory}}$, such as homogeneous plasma condition. Another noticeable feature is that the wave frequency $f_\textup{wave}$ seems to increase with $B_0$, but the wavelength remains relatively the same $\lambda_\textup{wave} \approx 1$ m, comparable to the radius of the chamber ($\sim 1.5$ m), suggesting that the wavelength could be bounded by the size of the plasma in BRB. In addition, we find that the waveforms shown in Figs.~\ref{fig:waveB}(a), \ref{fig:waveB}(d), and \ref{fig:waveB}(g) reverse their polarity when we flip the direction of the Helmholtz field ($\mathbf{B}_0 \rightarrow -\mathbf{B}_0$), which is consistent with the theoretical prediction from the induction equation in Eq.~(\ref{eq:MHD3}).

\begin{table*}
\caption{Summary of measured wave properties, such as wave frequency $f_\textup{wave}$, phase velocity $v_{p\text{-}\textup{exp}}$, and wavelength $\lambda_\textup{wave}=v_{p\text{-}\textup{exp}}/f_\textup{wave}$ at various $B_0$. The theoretical phase velocities $v_{p\text{-}\textup{theory}}=\sqrt{V_A^2+V_S^2}$ are also listed here for comparison, where $V_A$ and $V_S$ are the Alfvén speed and sound speed, respectively. The error ranges for the experimental values are $\pm 1\sigma$, resulting from uncertainty in the linear fitting. The speed $V_A$ and $V_S$ are estimated based on $T_e \approx 15$ eV and $n_e \approx n_i \approx 1 \times 10^{13}$ $\textup{cm}^{-3}$ from a previous measurement \cite{byvank_formation_2021}. Note that due to the nature of pulsed-power experiments, the actual $T_e$ and $n_e$ can vary significantly from shot to shot, leading to a large uncertainty in $v_{p\text{-}\textup{theory}}$. }
\label{tab:waves}
\begin{ruledtabular}
\begin{tabular}{cccccccc}
$B_0$ (G) & $I_\textup{bias}$ (A) & $f_\textup{wave}$ (kHz) & $\lambda_\textup{wave}$ (m) & $V_A$ (km/s) & $V_S$ (km/s) & $v_{p\text{-}\textup{exp}}$ (km/s) & $v_{p\text{-}\textup{theory}}$ (km/s)\\
\hline
20 & 3 & $29.6 \pm 0.5$ & ${\sim}1.0$ & ${\sim}14$ & ${\sim}38$ & $30.8 \pm 7.3$ & ${\sim}40$\\
40 & 3 & $35.9 \pm 4.3$ & ${\sim}0.7$ & ${\sim}28$ & ${\sim}38$ & $24.8 \pm 1.8$ & ${\sim}47$\\
60 & 3 & $53.6 \pm 2.5$ & ${\sim}1.1$ & ${\sim}42$ & ${\sim}38$ & $60.9 \pm 40.0$ & ${\sim}56$\\
60 & 0 & $41.3 \pm 1.9$ & ${\sim}0.9$ & ${\sim}42$ & ${\sim}38$ & $35.7 \pm 31.5$ & ${\sim}56$\\
\end{tabular}
\end{ruledtabular}
\end{table*}

Another prominent feature of the fast magnetosonic waves observed in the experiment is that they are clearly damped, as shown in Figs.~\ref{fig:waveB}(a), \ref{fig:waveB}(d), \ref{fig:waveB}(g), and \ref{fig:waveB}(j). We estimate the damping rate $\gamma$ based on the envelopes of the wave patterns and find that $\gamma \approx 3.7 \times 10^4$ $\textup{s}^{-1}$ or $\tau=1/\gamma \approx 27.1$ $\mu s$.

\subsection{Effects of the Poloidal Bias Flux $\Psi_\textup{gun}$}
\label{subsec:biasfluxgrid}




To study how the poloidal bias flux $\Psi_\textup{gun}$ in the CT injector affects the wave formation, we repeat the measurement in Figs.~\ref{fig:waveB}(g)--\ref{fig:waveB}(i) with $I_\textup{bias}=0$~A and the results are shown in Figs.~\ref{fig:waveB}(j)--\ref{fig:waveB}(l). The explanation of the data in each panel can be found in Sec.~\ref{subsec:HelmholtzB}. Note that the plasma accelerated out of the CT injector is unmagnetized when $I_\textup{bias}=0$ A. We can see in Table~\ref{tab:waves} that the properties of the waves driven by the unmagnetized plasma is comparable to those where $I_\textup{bias}=3$ A, indicating that it is primarily the plasma itself, rather than the CT embedded magnetic field that interacts with the background $\mathbf{B}_0$ and excites the fast magnetosonic modes in our experiment.

\section{Discussion}
\label{sec:discussion}

\subsection{Wave Driving Mechanisms}

Previous studies have shown that conducting spheres moving across magnetic field lines through a magnetized plasma experience an MHD drag force, causing the spheres to slow down and emit Alfvén and slow magnetosonic waves due to MHD Cherenkov radiation \cite{barnett_radiation_1986, parks_refueling_1988, newcomb_magnetohydrodynamic_1991}. This is a topic of importance with many applications to laboratory and space plasma physics, such as CTs injected into a tokamak plasma as a viable refueling scheme, or interaction of the Galilean satellites with the Jovian magnetosphere \cite{neubauer_sub-alfvenic_1998}. 


In our experiment, however, no MHD drag force is exerted on the CT since it is injected into the chamber along the magnetic field lines. One possible driver of the waves though is plasma expansion. We have shown in Sec.~\ref{sec:background} and Fig.~\ref{fig:wave}(b) that the fast magnetosonic mode observed in BRB is associated with non-zero perturbations in the flow velocity ($\mathbf{u}_1$) along the radial direction. Since the CT primarily has a cylindrical shape (radius $r \approx 30$ cm and length $l \approx 80$ cm) in the center of the chamber, the CT expansion leads to perturbations in the plasma flow along the $R$-axis, hence driving the fast magnetosonic waves in the experiment. 


After close examination of the magnetic topology measured by probe array MP2, we find that there are also small-amplitude fluctuations existing in the $R$- and $\phi$-direction, barely visible in Figs.~\ref{fig:idwave}(a)--\ref{fig:idwave}(b) and Figs.~\ref{fig:idwave}(g)--\ref{fig:idwave}(h). One possible explanation is that the wave vector of the fast magnetosonic mode $\mathbf{k}$ is not perfectly along the $R$-axis (i.e., the wave magnetic field $\mathbf{B}_1$ not perfectly parallel with the background field $\mathbf{B}_0$), causing perturbations in magnetic field in the $R$- and $\phi$-direction. Since the CT in our experiment has a non-uniform density profile, multiple MHD modes are expected to coexist \cite{saito_modes_2008}. Therefore, the small-amplitude fluctuations mentioned above can also be other MHD waves, such as shear-Alfvén waves or slow magnetosonic waves. Further experiments are required to identify the nature of these fluctuations. 

\subsection{Target Formation with Tangled Magnetic Fields for MIF}

Some potential methods have been proposed in the past to create the plasma target with tangled magnetic fields for MIF, such as injection of multiple gun-formed magnetized plasmas into a limited volume \cite{ryutov_adiabatic_2009, hsu_magnetized_2019}. In this experiment, we intended to randomize the magnetic field lines through turbulence resulting from colliding the CT plasma with the coarse grid in BRB. However, the magnetic field topology in the postgrid region presented in Figs.~\ref{fig:idwave}(a)--\ref{fig:idwave}(c) suggests otherwise. The magnetic field embedded in the CT is not perturbed by the grid; the CT mostly passes through the gird and excites the fast magnetosonic modes with oscillating magnetic fields along the $Z$-axis. 



In laser-driven turbulence experiments, it is shown that the Reynolds number (ratio of flow inertia to viscosity) is usually large enough that the effect of viscosity is negligible, resulting in a turbulent flow \cite{robey_time_2003, zhou_rayleightaylor_2017}. Inspired by these results, one can attempt to increase the flow Reynolds number so that transition to turbulence is more likely to occur when the CT plasma collides with the grid. Since the plasma kinematic viscosity $\eta \propto T^{5/2} \rho^{-1}$, where $T$ is the temperature and $\rho$ is the plasma mass density \cite{vold_plasma_2021}, higher Reynolds number might be achieved by decreasing the CT temperature and increasing the density and speed of the CT plasma.

\section{Summary}
\label{sec:summary}

In this paper, we present detailed experimental study of the fast magnetosonic waves driven by injection of compact toroid (CT) plasma along a magnetic field in BRB. By comparing the magnetic field topology obtained from a $\dot{B}$ probe array with MHD theory, our results indicate that the waves observed in BRB are the fast magnetosonic modes propagating perpendicular to the background magnetic field. Furthermore, we find that the wave frequency increases with the background field strength, but the wavelength (${\sim}1$ m) remains close to the radius of the chamber, suggesting that the wavelength is likely limited by the plasma size in BRB. By removing the preapplied poloidal magnetic flux in the CT injector, we show that it is primarily the plasma itself, not the field embedded in the CT that interacts with the background magnetic field and excites the fast magnetosonic modes. Last, our results suggest that the coarse conducting grid in the chamber, intended to randomize the magnetic fields embedded in the CT, has no effects on wave formation. Since part of this investigation is to form a target plasma with tangled magnetic fields for MIF, in future experiments, we propose to increase the Reynolds number in the CT so that transition to turbulence is more likely to occur during collision between the CT plasma and the grid. 


\section*{Supplementary Material}

See the supplementary material for an example of the magnetic field topology measured using the linear magnetic probe array MP1, which is located upstream from the grid as shown in Fig.~\ref{fig:exp}(b). 

\begin{acknowledgments}

Research presented in this paper was supported by the U.S. Department of Energy (DOE) Office of Fusion Energy Sciences through the Los Alamos National Laboratory under Contract No. 89233218CNA000001. Los Alamos National Laboratory is operated by Triad National Security, LLC, for the National Nuclear Security Administration of U.S. Department of Energy. The WiPPL User Facility is supported by the DOE Office of Science, Fusion Energy Sciences under Contract No. DE-SC0018266.


\end{acknowledgments}

\section*{Data Availability}

The data that support the findings of this study are available from the corresponding author upon reasonable request.

\bibliography{../refs}

\end{document}